# Negative index metamaterial combining magnetic resonators with metal films

Uday K. Chettiar, Alexander V. Kildishev, Thomas A. Klar[†], and Vladimir M. Shalaev

*Birck Nanotechnology Center, Purdue University, West Lafayette, IN 47907, USA*
[†]*on leave from Ludwig-Maximilians-Universität, München, 80799, Germany*
shalaev@purdue.edu

**Abstract:** We present simulation results of a design for negative index materials that uses magnetic resonators to provide negative permeability and metal film for negative permittivity. We also discuss the possibility of using semicontinuous metal films to achieve better manufacturability and enhanced impedance matching.

## 1. Introduction

The refractive index is a key parameter describing the interaction of light with matter. The real part of a refractive index is usually considered to be positive, which is in fact true for all naturally existing materials. However, a refractive index with a negative real part does not violate any physical laws; indeed such materials will have some very interesting properties making them good candidates for a plethora of valuable applications. Since such materials do not exist in nature they have to be artificially fabricated. In a material characterized by a permittivity $\varepsilon = \varepsilon' + i\varepsilon''$ and permeability $\mu = \mu' + i\mu''$, the condition $\varepsilon' < 0$ and $\mu' < 0$ is sufficient for negative refractive index [1]. But the condition $\varepsilon''\mu' + \mu''\varepsilon' < 0$ represents the necessary condition for a negative refractive index in a passive medium [2], where for example, a material with a large loss can have a negative refractive index even if the real parts of both permeability and permittivity are not negative simultaneously. Such a material will undoubtedly have a large loss preventing any useful application.

The first experimental demonstration of a negative index material (NIM) was given in the microwave frequency range using an array of metal lines and split ring resonators [3]. Negative index of refraction was demonstrated in optical frequencies using an array of paired nanorods [4], an array of elliptic voids in a multilayered structure [5] and a fishnet structure [6]. In this paper we present a simpler and intuitive 2D geometry which shows a pronounced negative refractive index.

## 2. Magnetic resonator based on a pair of metal nanostrips

It was predicted that a pair of metal rods separated by a dielectric can give rise to artificial permeability due to a localized plasmonic resonance [7-9]. In fact a pair of nanorods can have two kinds of resonance: a symmetric resonance, which results in an artificial permittivity and an asymmetric resonance, which gives rise to an artificial permeability; the two resonances make it possible to have a negative refractive index as was first predicted in [8]. In general, these two resonances occur at different wavelengths. We use a simplified 2D version of a nanorod pair as a magnetic resonator. This structure consists of two nanostrips separated by a dielectric as shown in Fig. 1(a). The strips are infinite in the direction perpendicular to the plane of the page (*y*-direction). The sample consists of an array of nanostrip pairs distributed periodically along the *x* direction. The incoming field is oriented in TM polarization as shown in Fig. 1(a). Similar to the nanorod pair, the nanostrip pair has two different principal resonances (the magnetic and electric resonance respectively). However, we are interested in using this structure exclusively as a magnetic resonator. The metal nanostrips are assumed to be made of silver because of its low losses. The dielectric spacer between the nanostrips is taken to be alumina because the relatively large refractive index of alumina is of assistance to confining the field between the nanostrips. The space between adjacent pairs of nanostrips is filled with silica to provide mechanical support for further lamellar stacking of this structure.

The complex transmission and reflection coefficients of this structure were calculated using a spatial harmonic analysis based code and verified with a commercial FEM solver. The permittivity of silver was taken from tabulated experimental values [10]. The transmission and reflection coefficients were then used to retrieve the effective refractive index and impedance for the nanostrip composite [11, 12]. Fig. 1(c) and (d) show the effective permeability and permittivity for the following design parameters, t = 30 nm, d = 40 nm, w = 300 nm, p = 600 nm (refer Fig. 1(a) for the key to the symbols). We note that there are two distinct resonances: a magnetic resonance around 1700 nm and an electric resonance around 900 nm. We also note the presence of an electric and a magnetic anti-resonance at 1700 nm and 900 nm, respectively. This is the result of the periodicity in the structure and has been studied extensively [13]. Because of the presence of the anti-resonances it is hard to overlap the magnetic and electric resonances, since as the electric and magnetic resonances get closer to each other the anti-resonances increase in strength, nullifying the resonance for one of the two quantities (permittivity or permeability).

The transmittance and reflectance characteristic is shown in Fig. 1(b). We can clearly see the relations between the Fig. 1(b) and Fig. 1(c,d). The real and imaginary parts of the permeability around 1700 nm show good correspondence with the transmittance and absorbance as expected. On the other hand the real part of the permittivity around 900 nm

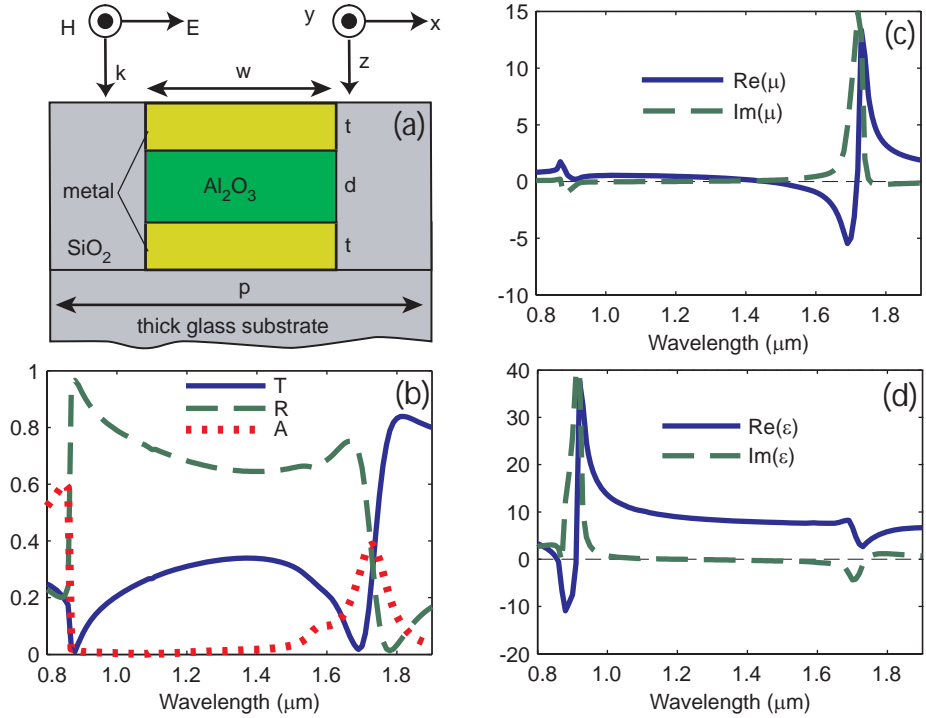

Fig. 1. (a) Unit cell for the array of nanostrip pairs; (b) Transmittance (T), reflectance (R) and absorbance (A) for the nanostrip pair array; (c) Effective permeability; (d) Effective permittivity

shows good correspondence with the transmittance, but there is no absorption corresponding to the electric resonance. There is an absorption band below 860 nm, but that corresponds to the radiation loss since the structure starts diffracting light in the forward direction below 860 nm. Hence the electric resonance does not result in any appreciable absorbance.

### 3. NIM using a magnetic resonator and continuous metallic films

In the previous section we saw that a pair of nanostrips can generate a strong magnetic resonance resulting in a negative permeability. Now all we need for creating a negative index material is a negative permittivity. This is easy to achieve since noble metals like gold and silver have negative permittivity at optical frequencies below the plasma frequency. Hence just adding a metal film above and below the magnetic resonator as shown in Fig. 2(a) should be sufficient to provide a negative permittivity. We note that an alternative way to provide negative permittivity is to use continuous wires [14]. In our design the films are brought in contact with the strips to avoid additional resonances due to the interaction between the strips and the films. Our simulations have shown that such additional resonances are detrimental to the magnetic properties of the nanostrip pair. The additional layer of silica on the top functions as a seal preventing the silver from being oxidized or sulfurized due to exposure to air. Intuitively, only one metal film should suffice since we can use a single thick film instead of two thin films to provide similar permittivity. There are two reasons for using two films instead of one. Using two films ensures that the structure is symmetric and justifies the use of the retrieval formula to extract the effective optical parameters of the structure. The retrieval procedure assumes that the structure is symmetric. Using a single film would also generate additional resonances due to the interaction between the isolated strip and the film. This interaction is suppressed through the use of two films.

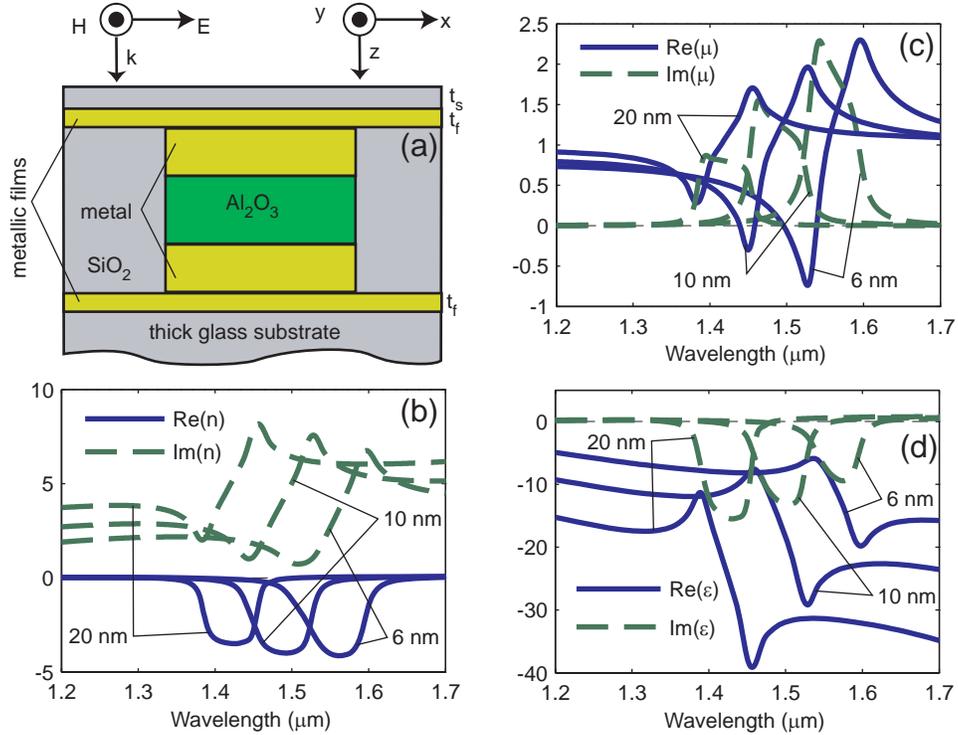

Fig. 2. (a) Unit cell with a magnetic resonator and metallic (silver) films on top and bottom; (b) Effective refractive index with $t_f$ = 6 nm, 10 nm and 20 nm. For all cases $t_s$ = 20 nm; (c) Effective permeability; (d) Effective permittivity.

The dimensions of the nanostrips were taken from the previous section and simulated for 6 nm, 10 nm and 20 nm silver films. The structure was coated with a 20 nm thick silica layer. The effective refractive index, permeability and permittivity are shown in Fig. 2. Here the refractive index goes negative for all cases, but the permeability goes negative only when the metal film is 6 nm or 10 nm thick. Also the imaginary part of the refractive index is much lower when the real part of the permeability is negative. The sample with 6 nm thick films has a maximum transmittance of 27% with a negative real part of the refractive index at a wavelength of 1520 nm. The refractive index at this wavelength is −2.38+0.84i. However it should be remembered that it is almost impossible to fabricate a 6 nm thick continuous silver film. The silver film would be essentially semi-continuous at this thickness. Typically, the film thickness should be at least 20 nm to ensure that the silver is continuous, but as it is clear from Fig. 2, the properties are significantly worse with a 20 nm thick film. For example the minimum of $\mathrm{Im}(n)$ is about 2.5 times as large for the 20 nm thick silver films as compared to the 6 nm thick films (Fig. 2(b)). This is a direct consequence of $\mathrm{Re}(\mu) > 0$ (Fig. 2(c)).

We also notice that the magnetic resonance is at shorter wavelengths for the thicker films. The addition of the metal films causes the currents in the nanostrip to leak into the metal films, especially towards the ends of the nanostrips. This reduces the effective width (w) of the nanostrips which results in a shorter resonance wavelength. This leakage of current also has the effect of diminishing the resonance in permeability.

### 4. NIM using a magnetic resonator and semicontinuous metallic films

So far, our simulations have shown that combining continuous metal film with nanostrip magnetic resonators could easily yield a negative index material. Unfortunately, that requires unrealistically thin metal films. The problem arises because silver has a highly negative real part for it's permittivity around 1.5 μm, but the magnetic resonator can not provide a comparable value of negative permeability. This results in a huge impedance mismatch and subsequent large reflection which degrades the magnetism of the entire structure. Another way to look at the problem is to consider the electromagnetic shielding provided by the metal film. A thick film would shield the incoming field more efficiently, thus suppressing the performance of the magnetic resonator. An obvious way to circumvent this problem is to use a thinner metal film, except the minimum thickness is limited by the material properties and fabrication tolerances.

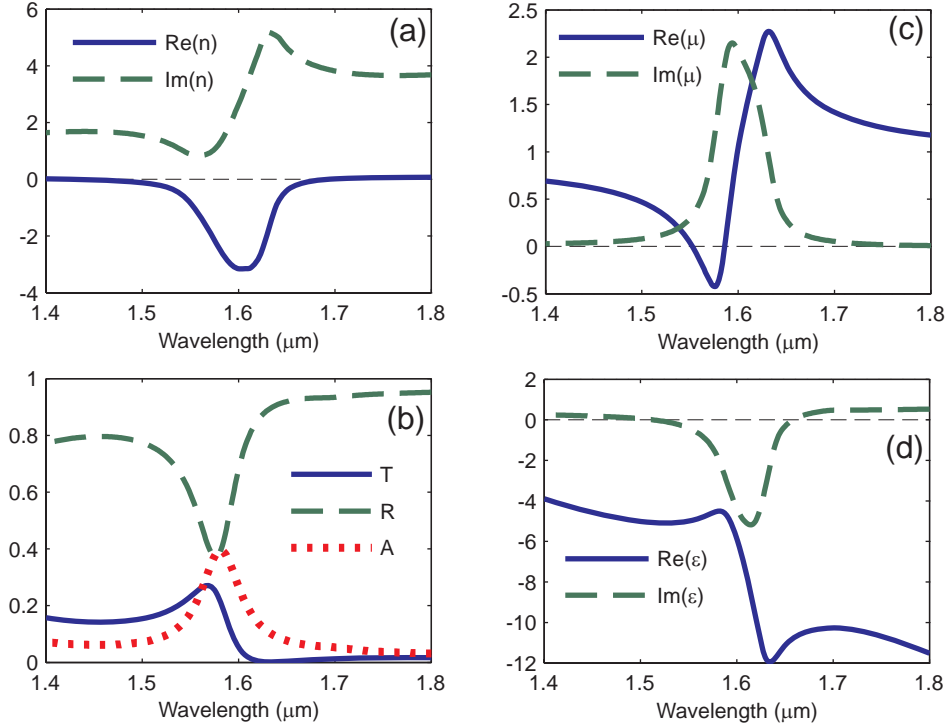

Fig. 3. Effective parameters and spectrum for the structure with semicontinuous silver film (metal filling fraction = 65%, thickness = 20 nm). (a) Effective refractive index, (b) Transmittance, reflectance and absorbance spectra, (c) Effective permeability, (d) Effective permittivity.

Instead of bulk metal, the film can be formed with a mixture of silica and metal (semicontinuous metal films, SMF). Such films can be fabricated using very basic techniques like sub-monolayer evaporation. The permittivity of a SMF can be well described by the effective medium theory (EMT) [15]. According to EMT, the effective permittivity ($\varepsilon_e$) of a $d$-dimensional metal-dielectric composite consisting of a metal with permittivity $\varepsilon_m$ and a filling fraction of $f$, and a dielectric with permittivity $\varepsilon_d$ and filling fraction of $1-f$ is given by Eq. (1). This is a quadratic equation which has two solutions for $\varepsilon_e$. We select the solution that has a positive imaginary part ($\mathrm{Im}(\varepsilon_e) \geq 0$).

$$f \frac{\varepsilon_m - \varepsilon_e}{\varepsilon_m + (d-1)\varepsilon_e} + (1-f) \frac{\varepsilon_d - \varepsilon_e}{\varepsilon_d + (d-1)\varepsilon_e} = 0 \tag{1}$$

The performance of EMT has been tested against exact solution using numerical methods [16] and it has been confirmed that EMT provides good agreement with the numerical method except for some minor differences. According to EMT, the permittivity of a SMF is much less negative as compared to bulk metal films. Hence we have a SMF that is thick enough to be fabricated easily while having a real part of the permittivity which can be matched with the permeability created by magnetic resonators. This is critical for impedance matching.

For example, consider a structure similar to that of Fig. 2, simulated in the previous section. This time, semicontinuous silver films are used instead of the continuous silver films. We assume the SMF to have a metal filling fraction of 65%. The physical thickness of the SMF is 20 nm, which is quite easy to achieve. The permittivity of the SMF was calculated using two dimensional EMT. The results of the simulation are shown in Fig. 3.

The real part of the refractive index is negative between 1500 nm and 1650 nm. The transmittance spectrum has a maximum value of 27% at 1570 nm. The refractive index has a value of −1.85+0.93i at the wavelength of maximum transmittance. We note that the transmittance is better than the case with 10 nm continuous silver film although the SMF contains more metal than a 10 nm continuous silver film.

## 5. Conclusions

We examined the magnetic behavior of nanostrip pair arrays. The nanostrip pair array was found to posses a strong magnetic resonance, which results in a negative permeability. By changing the dimensions of the nanostrip, the region of negative permeability is easy to tune over a wide range of wavelengths.

The nanostrip pair array can be modified without difficulty to exhibit a negative refractive index by fusing metal films on the top and bottom of the nanostrip pair. The fusing of the films with the nanostrips is essential to avoid additional resonances due to the interaction of the nanostrips with the films.

A homogeneous silver film provides a highly negative permittivity, which can not be matched with the permeability provided by the magnetic resonator. The resulting impedance mismatch significantly deteriorates the properties of the complete structure. Hence we use semicontinuous metal films (SMF) to take this problem into account. The use of SMFs provides an additional degree of freedom in optimizing the behavior of such metamaterials.

Certainly, there are other ways to achieving the same goal, using for example an array of elliptical voids [5] or a fishnet geometry [6]. The fishnet geometry is typically fabricated using E-beam lithography, which severely limits the size of the sample. In contrast, a larger array of elliptical voids can be fabricated with interference lithography, though such a sample has the possible disadvantage of variable width of the magnetic resonators, since both the shape and dimensions of the voids are limited by the resolution of the fabrication process. In our geometry a large array of magnetic resonators can be fabricated with a high fidelity using interference lithography, since only two interfering beams are required for fabricating the magnetic substructure. The semicontinuous films can be deposited efficiently using a standard E-beam or thermal evaporation. Thus our strategy allows for the fabrication of larger structures through relatively inexpensive means.

We could also extend the idea and fabricate the nanostrips using semicontinuous metal instead of bulk metal. This will provide us with more flexible means of making manufacturable resonant metallic elements and adjusting their properties.

## Acknowledgements

This work was supported in part by NSF-NIRT award ECS-0210445, by ARO grant W911NF-04-1-0350 and by ARO-MURI award. TAK wants to thank the Alexander von Humboldt Foundation for a Feodor-Lynen-Stipendship. We would like to acknowledge fruitful collaboration with Wenshan Cai and the rest of the members of the Photonics and Spectroscopy Lab at Birck Nanotechnology Center, Purdue University.